\documentstyle[epsfig]{aipproc}

\begin{document}
\renewcommand{\thefootnote}{\fnsymbol{footnote}}

\title{Results from \boldmath $\gamma \gamma$ \unboldmath collisions in
 OPAL\footnotemark[1]
}

\author{Jochen Patt}
\address{Fakult\"at f\"ur Physik, Albert-Ludwigs-Universit\"at Freiburg, Germany}

\maketitle

\vspace{-0.5cm}

\begin{abstract}
The production of charged hadrons and jets is measured in collisions
of quasi-real photons. The data were taken 
with the OPAL detector at LEP at e$^{+}$e$^{-}$ centre-of-mass energies
$ \rm \sqrt{s_{ee}}$ = 161 and 172 GeV. The measured cross-sections are compared to
perturbative next-to-leading order QCD calculations. 
The separation of the direct and the resolved component of 
the photon is demonstrated.  
\end{abstract}

\vspace{-0.5cm}

\section*{Introduction}
\setcounter{footnote}{0}
\renewcommand{\thefootnote}{\fnsymbol{footnote}}
\footnotetext[1]{Talk given at the 
7th International Conference on Hadron Spectroscopy (HADRON 97),
Brookhaven, 25-30 August 1997, Freiburg preprint EHEP-97-18}

In an e$^+$e$^-$ collider both beam electrons can
radiate virtual photons which interact with each other. 
These photons 
are described by their negative four-momentum transfer $Q^2$
which is usually small ($Q^2 \approx 0$). They can therefore
be considered to be quasi-real. Accordingly the beam electrons are scattered with
very small angles and are not detected (anti-tagged). Events where one or 
both scattered
electrons are detected are called single-tagged or double-tagged~[1].\\
The interactions of the photons can be modelled by assuming that each photon can
either interact directly or appear resolved through its fluctuations into
hadronic states. In leading order Quantum Chromodynamics (QCD) this model leads to
three different event classes for $\gamma \gamma$ interactions: direct, single-resolved
and double-resolved. In resolved events the partons (quarks or gluons) inside 
the photon take part in the hard interaction. The probability to find partons
inside the photon is parametrised by parton density functions.\\
Due to the direct photon interactions the transverse momentum 
distribution of the charged hadrons in two-photon
interactions is expected to have a harder component than in photon-proton
or hadron-proton interactions. This is
demonstrated by comparing the $\gamma \gamma$ data to photo- and hadroproduction data measured
by WA69~[2].\\
Events with two jets (dijet events) offer the possibility to separate
direct and resolved processes. The jet cross-sections can be compared to
QCD calculations using different structure function parametrisations.
Resolved processes are especially sensitive to the gluon 
content of the photon.

\section*{Hadron production in \boldmath $\gamma \gamma$ \unboldmath events}

We present the first measurement of inclusive charged hadron production in anti-tagged
$\gamma \gamma$ collisions. The data were taken in 1996 and
correspond to an integrated luminosity of 20 pb$^{-1}$.\\
To select anti-tagged two-photon events a set of cuts is applied to the data,
the most important cuts are: The visible invariant hadronic
mass $W_{\rm ECAL}$ measured in the electromagnetic calorimeter has to be greater than 3 GeV, to reject multihadronic
events the sum of all energy deposits in the electromagnetic and hadronic calorimeters should be less
than 45 GeV and at least 3 tracks must have been found in the tracking chambers which rejects lepton pair
events. To select only anti-tagged events there is a cut on the maximum energy deposit in the
forward and silicon tungsten calorimeters that are located at small angles. After applying the whole set of cuts around 59000 events
remain.   
\begin{figure}[htbp]
\vspace*{-20pt}
\centerline{\epsfig{file=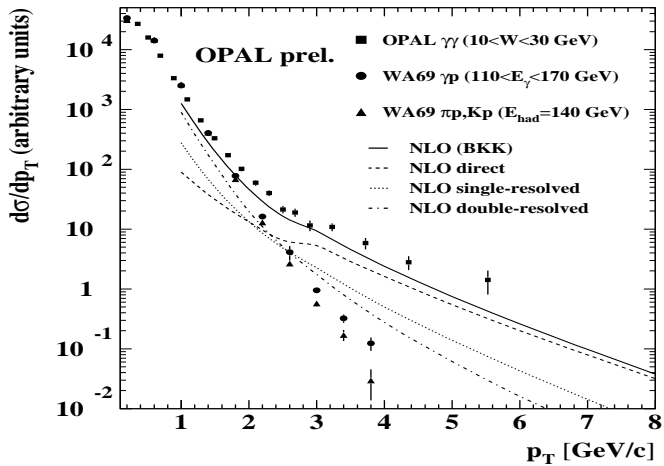,width=4.0in,height=3.0in}}
\caption{\label{F:jpatt:1} 
 The distribution  measured in $\gamma \gamma$
 interactions is compared to the
 $p_{\rm T}$ distribution measured in 
 $\gamma$p and 
 hp (h=$\pi$,K) interactions in WA69.}
\end{figure}
Fig.~\ref{F:jpatt:1} shows the differential cross-section d$\sigma$/d$p_{\rm T}$ as a function of the transverse
momentum $p_{\rm T}$ of the hadrons. The data are compared with results from $\gamma$p and hp scattering
(h=$\pi$,K) from the WA69 experiment which have been normalised to the $\gamma \gamma$ data.
At $p_{\rm T} >$ 2 GeV/c the data lie significantly above the exponentially falling $\gamma$p and hp data
due to the contribution from the direct $\gamma \gamma$ interactions. 
The $\gamma \gamma$ data are also compared to next-to-leading order (NLO) 
calculations~[3] containing the three components of the direct,
single-resolved and double-resolved processes. The calculations show clearly the dominance of the direct 
component at large transverse momenta.
Most hadrons have small transverse momenta, in a regime where 
the resolved component dominates. This means that in  
most cases the photon behaves like a hadron.

\section*{Jet production in \boldmath $\gamma \gamma$ \unboldmath events}

\begin{figure}[htpb] 
\unitlength1cm
\begin{picture}(0.,5.5)
\put(0.,0.){
          \epsfxsize=5.5cm
          \epsffile{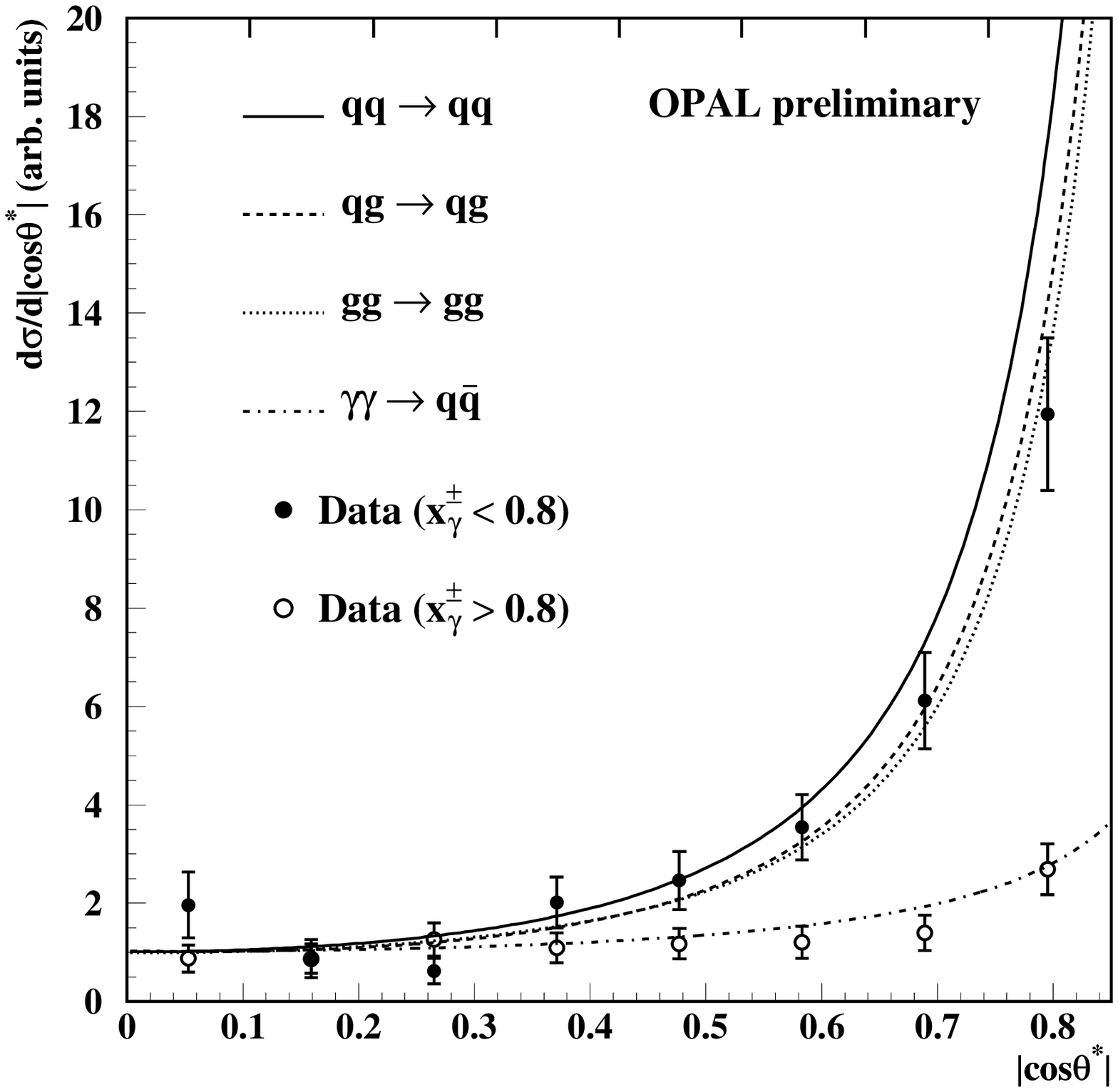}
          }
\put(7.,-0.3){
           \epsfxsize=6.cm
           \epsffile{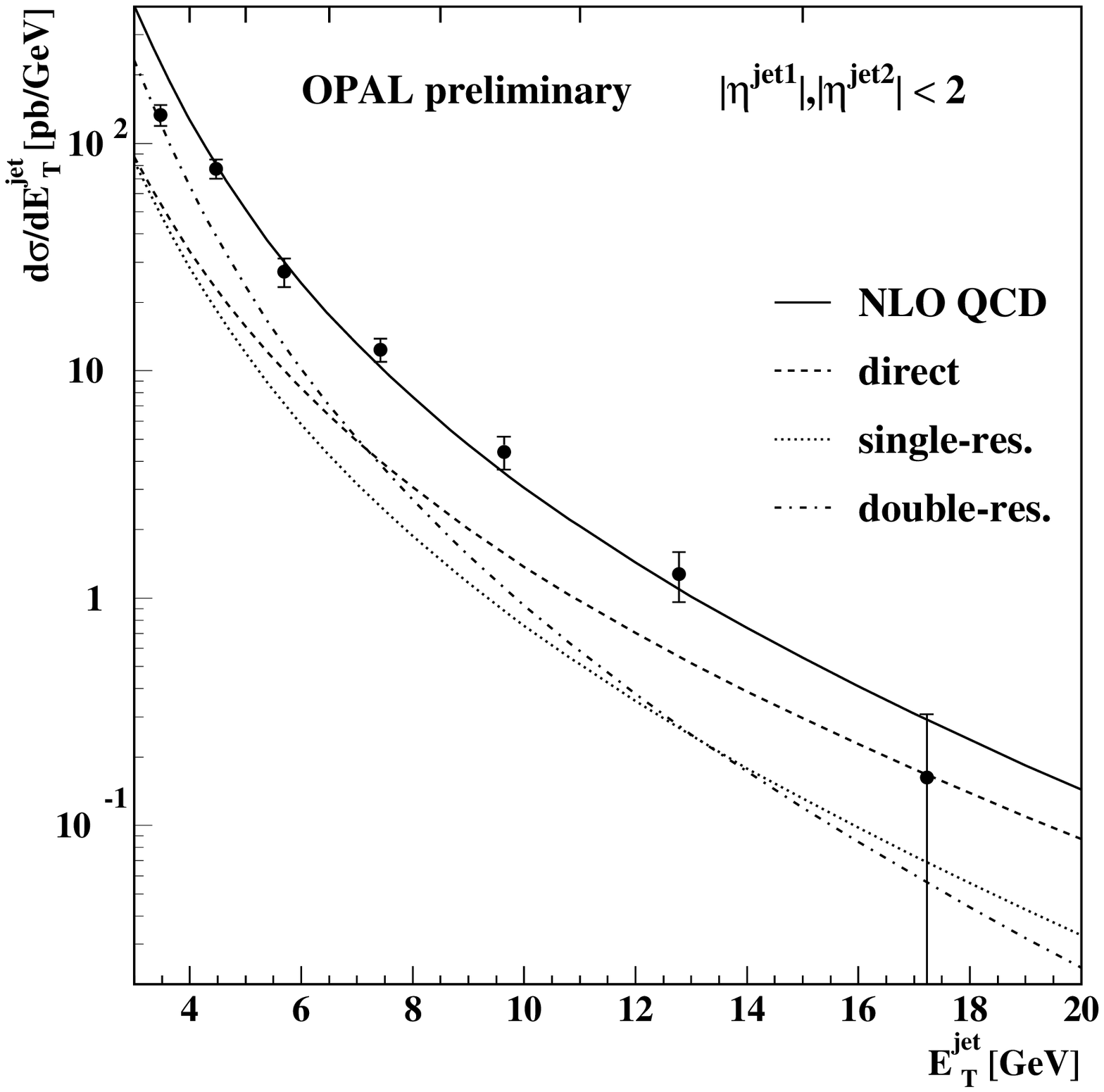}
           }
\end{picture}
\vspace*{10pt}
\caption{Angular distribution of
         events with large direct and large
         double-resolved components according to the separation with
         $x^+_{\gamma}$ and $x^-_{\gamma}$. The right plot shows the 
         inclusive two-jet cross-section as a function of $E^{\rm jet}_{\rm T}$.}
\label{F:jpatt:2}
\end{figure}

The same data and a similar set of cuts as in the charged hadron analysis are used to investigate the
production of jets in anti-tagged $\gamma \gamma$ interactions~[4]. To select jets a cone jet finding algorithm 
is applied to the data 
requiring a minimum transverse energy of the jets of $E^{\rm jet}_{\rm T}> $ 3 GeV, 
the pseudorapidity $ | \eta^{\rm jet} | $ to be less than 2 and the cone size $R$ to be equal to 1.\\
Especially events where two jets are found offer the possibility to 
differentiate between direct and resolved interactions~[5].
A pair of variables, $x^+_{\gamma}$ and $x^-_{\gamma}$, can be defined which is a measure of the
photon's energy participating in the hard interaction:
$ x^{\pm}_{\gamma}={\sum_{\rm jets}(E \pm  p_z)}/{\sum_{\rm had}(E \pm p_z)}$,
where $p_{\rm z}$ is the momentum component along the $z$ axis of the detector and
$E$ is the energy of the jets or hadrons.
In direct events the whole energy is concentrated in the two jets
whereas in resolved events there is also energy outside the two jet cones due to
the remnant of the resolved photon(s). 
So events with $x^{\pm}_{\gamma} >$ 0.8 mainly stem from direct and
events with $x^{\pm}_{\gamma} <$ 0.8 mainly from resolved processes.\\
In direct events quark exchange dominates the interaction, whereas  
gluon exchange dominates in resolved events.  This gives rise to different angular distributions which can be seen
in Fig.~\ref{F:jpatt:2}. The parton (jet) scattering angle  
$| \cos\theta^{\ast}|$ in the centre-of-mass system of the outgoing partons (jets) is plotted. 
The two components are well separated, in the data as well as in
the calculations.
The right plot in Fig.~\ref{F:jpatt:2} shows the inclusive two-jet
cross-section as a function of $E^{\rm jet}_{\rm T}$ for jets with $|\eta^{\rm jet}| < 2$ 
compared to NLO calculations by Kleinwort and Kramer~[6].
The agreement between the data and the
calculations is good apart from the lowest $E^{\rm jet}_{\rm T}$ bin. 
\vspace*{-5pt}

\begin{figure}[htpb]
\vspace*{-10pt}
\centerline{\epsfig{file=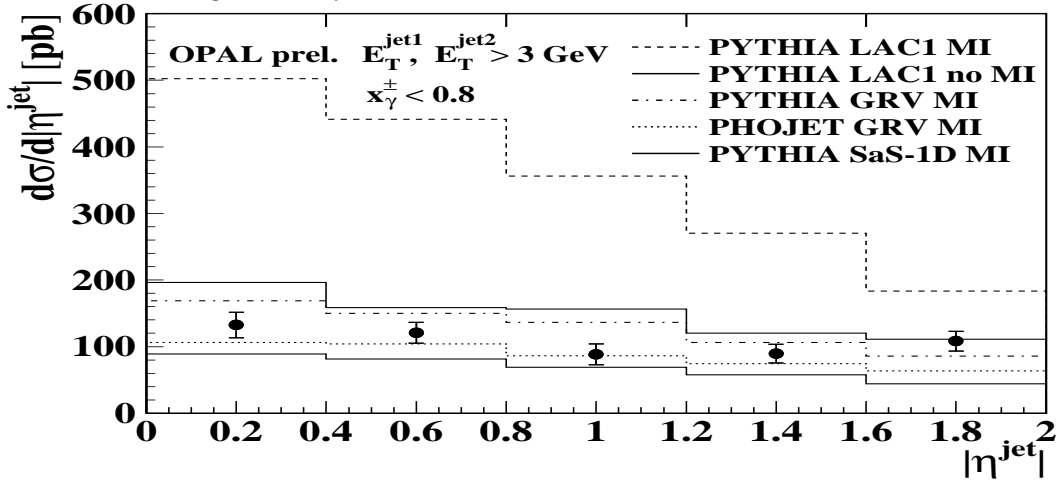,width=5.5in,height=2.5in}}
\caption{The inclusive two-jet cross-section as a function of $|\eta^{\rm jet}|$
for jets with $E^{\rm jet}_{\rm T} >$ 3 GeV for events with a large contribution
of double-resolved events (requiring $x_{\gamma}^{\pm} <$ 0.8).}
\label{F:jpatt:3}
\end{figure}

Especially double-resolved events are sensitive to the gluon content of the photon
described by different parton density functions. This leads to different
predictions of the inclusive two-photon cross-section. Fig.~\ref{F:jpatt:3}  
shows the inclusive two-photon cross-section as a function of $|\eta^{\rm jet}|$
for 
double-resolved events ($x_{\gamma}^{\pm} <$ 0.8) in comparison
with Monte Carlo models with different photon structure functions. The LAC1 
parametrisation~[7] overestimates the inclusive two-jet cross-section
significantly.

\vspace*{-20pt}
\section*{Conclusion}

The differential cross-section d$\sigma$/d$p_{\rm T}$ of charged hadrons in $\gamma \gamma$ events
was found to have a harder component  than in  $\gamma$p or hp scattering due to the presence of
the direct photon-quark coupling. The NLO calculations confirm the higher cross-section 
at large transverse momenta.\\
Events with two jets were identified using a cone jet algorithm. A pair of variables
$x^+_{\gamma}$ and $x^-_{\gamma}$ is used to separate experimentally the events emerging from
direct and resolved interactions. Both components are well separated in the distribution
of the parton (jet) scattering angle $| \cos\theta^{\ast}|$. The agreement between the data and
the NLO calculations is good.

\vspace*{-10pt} 

\end{document}